\documentclass[reprint,prl,nofootinbib,superscriptaddress]{revtex4-1}
\pdfoutput=1
\usepackage{graphicx,amsmath}
\newcommand{\eq}[1]{\begin{equation}#1\end{equation}}

\newcommand{\fig}[4]{\begin{figure}[#4]\centering\includegraphics[width=#3\textwidth]{Graph-#1.pdf}\caption{#2}\label{fig-#1}\end{figure}}

\newcommand{\refeq}[1]{eq.\ (\ref{eq-#1})}

\newcommand{\refig}[1]{FIG.\ \ref{fig-#1}}

\newcommand{\subs}[1]{_\mathrm{#1}}

\newcommand{\dd}[1]{\mathrm{d}#1}
\def\mpl{M\subs{p}}

\def\fNL{f\subs{NL}}

\newcommand{\cm}[1]{}
\def\fin{\phi}
\def\fcu{\sigma}
\def\Vi{V}
\def\Vc{U}
\def\Va{V\subs{tot}}

\def\rhoc{\rho_\fcu}

\def\st{c}
\def\sh{*}
\def\se{e}

\def\sst{{}_{\st}{}}
\def\ssh{{}_{\sh}{}}
\def\sse{{}_{\se}{}}

\begin{document}
\title{Creating perturbations from a decaying field during inflation}
\author{Anupam Mazumdar}
\affiliation{Consortium for Fundamental Physics, Physics Department, Lancaster University, LA1 4YB, UK}
\affiliation{Niels Bohr Institute, Copenhagen University, Blegdamsvej-17, Denmark}
\author{Lingfei Wang}
\affiliation{Consortium for Fundamental Physics, Physics Department, Lancaster University, LA1 4YB, UK}
\begin{abstract}
Typically the fluctuations generated from a decaying field during inflation do not contribute to the large scale structures. In this paper we provide an example where it is possible for a field which slowly rolls and then decays during inflation to create all the matter perturbations with a slightly red-tilted spectral index, with no isocurvature perturbations, and with a possibility of a departure from Gaussian fluctuations.
\end{abstract}
\maketitle

Primordial inflation is one of the most successful paradigms for the standard model cosmology which has many observational consequences~\cite{WMAP}. In spite of this success, understanding the origin of inflation is still very challenging~\cite{Infl-rev}. In the simplest version it is assumed that a light scalar field is slowly rolling down the potential which produces almost scale invariant Gaussian fluctuations~\cite{Brandenberger}.

However there could be departures from this simplest paradigm. In principle there could also be many scalar fields which might participate during inflation~\cite{assisted}, out of which some could be heavy and they might decouple from the dynamics early on, while leaving the lightest field to slow roll to end inflation, or some could be ultra light which might even decay after the end of inflation, such as in the case of a curvaton scenario~\cite{curvaton}.

One such physical possibility which we wish to explore here is the fate of a heavy scalar field during inflation. Since the field is very heavy compared to the Hubble expansion rate, one might expect it to settle down in its potential within one Hubble time. Its perturbations would be damped and as a result it would not leave any observable imprint on large scales, see for instance~\cite{Liddle}.

However this might not be the case if the heavy field lies initially on a plateau that is flat enough to accommodate slow roll, such as in the vicinity of a saddle or inflection point, very similar to the cases of inflation discussed in Refs.~\cite{MSSM}. If such a field rolls off the plateau during the last $50-60$ e-folds, then it is possible to seed the primordial fluctuations for the observable scales which will be dominated by the fluctuations of the heavy field while making the inflaton's own fluctuations redundant. 

In order to probe such a scenario, we need to understand the dynamics of both the fields, i.e. the heavy and the inflaton. Let us consider an inflaton $\fin$ with potential $\Vi(\fin)$, and another heavy field $\fcu$ with potential $\Vc(\fcu)$ added upon it, so the total potential is
\eq{\Va(\fin,\fcu)=\Vi(\fin)+\Vc(\fcu).}
We consider the case where the inflaton $\fin$ leads inflation, so $\Vi>\Vc$ always holds, whereas perturbations of $\fin$ are negligible compared to that of $\fcu$. We also assume $\Vc(\fcu)$ to have a plateau very flat and smooth to accommodate slow roll. To simplify the calculations we may assume that the $\Vc(\fcu)$ potential turns down very sharply at the plateau edge.

\fig{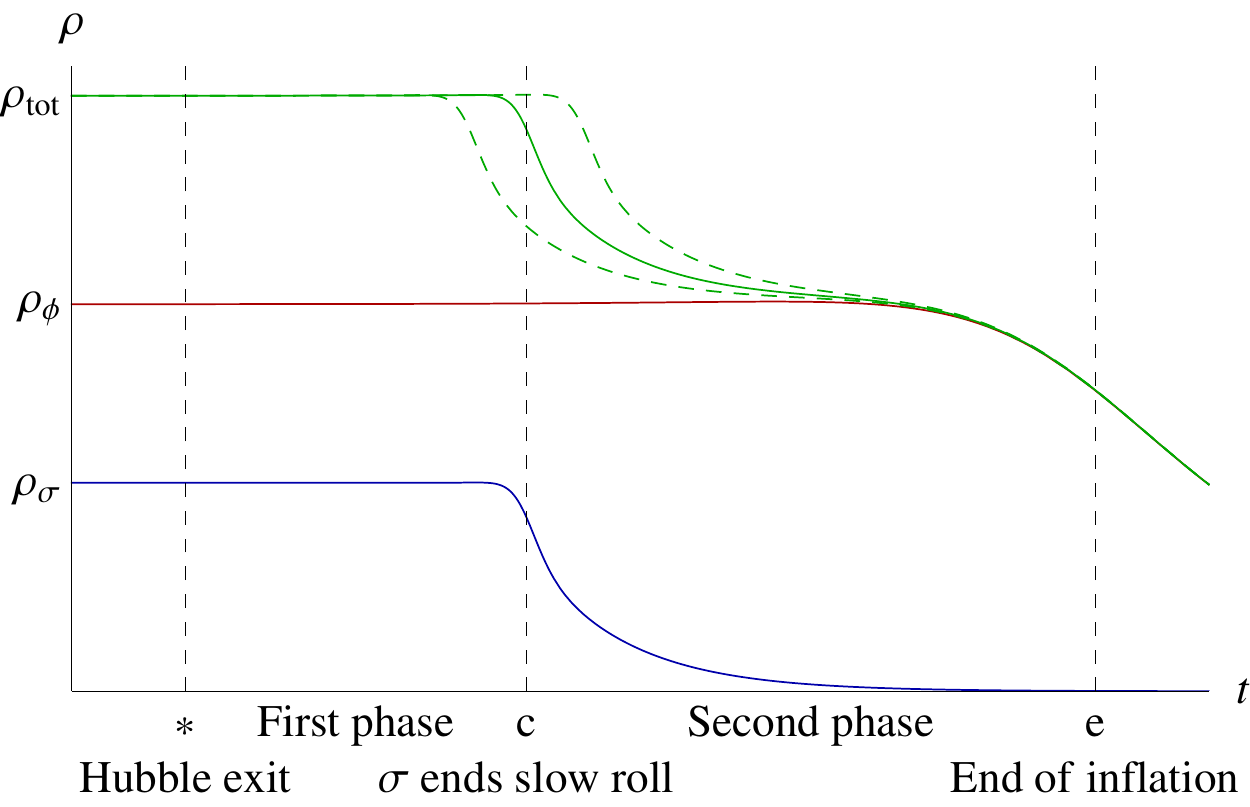}{The schematic timelines of background and perturbed universes. The blue, red and green solid lines denote $\fcu$, $\fin$ and total energy densities of the background universe respectively. The dashed green lines show the total energy densities in the perturbed universes where the main contribution comes from fluctuations in the $\fcu$ field. After the Hubble exit, inflation is divided into two phases by the break of $\fcu$'s slow roll condition.}{0.50}{}
Because of the plateau in $\Vc(\fcu)$, we can split inflation into two distinct phases with three critical points, whose timelines are shown in \refig{Timeline}. During the first phase, both the fields are slowly rolling as $\fcu$ remains on the plateau. This is terminated by the break of the second order slow roll condition of $\fcu$ field as it reaches the edge of the plateau. Then $\fcu$ rolls quickly to its minimum of the potential and oscillates, which we regard here as the second phase. During this phase $\fcu$ oscillates many times within one Hubble time and its quanta are quickly redshifted away by inflation. Several e-folds after $\fcu$ ends slow roll, it is already diluted enough so the universe enters a single field inflation.

As shown in \refig{Timeline}, perturbations in $\sigma$ field is actually responsible for a fluctuating boundary between the phases, generating different Hubble rates around the time $\fcu$ ends slow roll. This imparts fluctuations in the number of e-folds:
\eq{\delta N=N_{\fcu}\delta\fcu\ssh,}
where $N_{\fcu}\equiv \partial N/\partial\sigma\ssh$, and ``$\sh$'' indicates at the time of Hubble exit. The power spectrum for curvature perturbation is given by:
\eq{P_\zeta=P_{\delta N}=N_{\fcu}^2P_{\delta\fcu\ssh}\,,\label{eq-e-Pz0}}
in which $P_{\delta\fcu\ssh}=H\ssh^2/4\pi^2$ for any sufficiently flat potential during inflation, with $H_{\ast}$ being the Hubble rate of expansion when the relevant modes for observations  are leaving the Hubble patch.

Exact analytical calculations exist for two-field inflations \cite{twofield}, but we can estimate $P_\zeta$ easily by using the evolution of the inflaton as a clock.  As shown by the dashed lines in \refig{Timeline}, we see that the fluctuations in $\fcu$ lead to different locations for the end of slow roll condition in perturbed/background universes. As an estimation, we may assume that once $\fcu$ is off the plateau it doesn't contribute at all to the energy density. This simplifies $\rhoc$ to almost a step function instead of the one shown in \refig{Timeline}. Then the initial perturbation $\delta\fcu\ssh$ moves the location of $``c "$, (which indicates when $\fcu$ ends slow roll) by the amount of time
\eq{\delta t\approx-\delta\fcu\ssh/\dot\fcu\ssh=3H\sst\delta\fcu\ssh/\Vc\ssh'\,,\label{eq-e-dt}}
or by the amount of inflaton clock time:
\eq{\delta\fin\sst=\dot\fin\sst\delta t\approx-\frac{\Vi\sst'}{\Vc\ssh'}\delta\fcu\ssh\,.}
 During this time period in a perturbed universe, $\fcu$ would still contribute to the total energy density, while it would not in the background universe. This brings about the difference in the Hubble rate by:
\eq{\left.\frac{H_{pert}^2-H^2}{H^2}\right|_{\st}\approx\left.\frac{\Vc}{\Vi+\Vc}\right|_{\st}\,.\label{eq-e-dH}}

From the slow roll approximations we get the numbers of e-folds of inflation that are generated during this inflaton clock time $\delta\fin\sst$, for the background and the perturbed universes
\eq{N\approx-\frac{3H^2}{\Vi\sst'}\delta\fin\sst,\hspace{0.35in}N_{pert}\approx-\frac{3H_{pert}^2}{\Vi\sst'}\delta\fin\sst\,.}
Since the only difference between the perturbed and the background universes' evolutions comes from the period $\delta\fin\sst$, we arrive at\footnote{We have approximated $\Vc\sst \approx \Vc\ssh$ for a flat enough plateau.}
\eq{\delta N\equiv N_{pert}-N\approx\frac{8\pi\Vc\ssh}{\mpl^2\Vc\ssh'}\delta\fcu\ssh\,,\label{eq-e0-dN}}
so that
\eq{P_\zeta\approx\frac{16H\ssh^2\Vc\ssh^2}{\mpl^4\Vc\ssh'{}^2}\,.\label{eq-e0-Pz}}

From \refeq{e0-Pz}, we see that the result we have estimated for this scenario is very simple.  The spectrum is {\it solely}
governed by the  dynamics of $U(\sigma)$ instead of $V(\phi)$, and the spectral tilt of the curvature perturbations is dominated by:
\eq{n_s-1\equiv\frac{\dd\ln P_\zeta}{H\dd t}\approx-2\epsilon\ssh\,,\label{eq-e0-ns}}
where
\eq{\epsilon\ssh\equiv\frac{\dd}{\dd t}\frac{1}{H\ssh}}
is the total first order slow roll parameter. From \refeq{e0-ns} we achieve the red-tilt which is favored by current observations, i.e. the central value of the spectral tilt is given by: $n_s\approx 0.96$~\cite{WMAP}. Furthermore, since inflation goes on, $n_s-1$ will be further negative, the running in the spectral tilt will also be slightly negative, which is also favored by the current observations.\footnote{For the scales leaving Hubble patches after $\fcu$ ends slow roll, the perturbations from $\sigma$ field will be very suppressed. The relevant perturbations from the fluctuating $\sigma$ field must leave the Hubble patch before the phase boundary. This does not pose any strong constraint on $N\sst$ which is the number of e-folds from the phase boundary (denoted by $\st$) till the end of inflation (denoted by $\se$), see \refig{Timeline}.}

After providing the readers with this heuristic argument, let us now proceed with a detailed calculation which will also yield to 
the bispectrum of curvature perturbations. Starting from the first phase, the slow roll approximations give the equations of motion for both fields
\eq{\dd\fin/\dd t=-\Vi'/3H,\hspace{0.4in}\dd\fcu/\dd t=-\Vc'/3H\,,}
whose ratio after integration is
\eq{\int_{\fin\sst}^{\fin\ssh}\frac{\dd\fin}{\Vi'}=\int_{\fcu\sst}^{\fcu\ssh}\frac{\dd\fcu}{\Vc'}\,.\label{eq-e-slint}}

When being perturbed by the initial $\delta\fcu\ssh$, because $\fcu$'s plateau edge is sharp it will always end slow roll at almost the same position, regardless of the perturbation. This allows us to neglect $\delta\fcu\sst$ to get the perturbed version of \refeq{e-slint}:
\eq{\frac{\delta\fin\sst}{\Vi\sst'}+\frac{\delta\fcu\ssh}{\Vc\ssh'}=0\,.}
Then it is easy to write down the perturbed number of e-folds during the first phase, for a flat enough plateau
\eq{\delta N_1=-\frac{8\pi(\Vi\sst+\Vc\sst)}{\mpl^2\Vi\sst'}\delta\fin\sst=\frac{8\pi(\Vi\sst+\Vc\sst)}{\mpl^2\Vc\ssh'}\delta\fcu\ssh\,.\label{eq-e-dN1}}

After $\fcu$ ends slow roll, it may either oscillate around its vacuum with a frequency much higher than the Hubble rate, or it may decay instantly. Irrespective of these two scenarios we will have more or less a constant overall equation of state, $w$, for the $\sigma$ oscillations or its decay products. In either case the decay process is generally unimportant in this scenario as everything will be diluted away by the follow-up inflation.

Let us consider the second phase which starts as $\fcu$ ends slow roll. From the Friedmann equation and the equation of motion for the slow roll field $\fin$, we get
\eq{H^2=\frac{8\pi}{3\mpl^2}(\Vc\sst e^{-3(1+w)N}+\Vi)=\frac{\Vi'}{3\fin'}\,,}
where $\fin'\equiv\dd\fin/\dd N=-\dd\fin/H\dd t$. This equation is a first order differential equation between $\fin$ and $N$, which has an exact solution for the number of e-folds of the second phase:
\eq{N_2=n(\fin\sst,\fin\sse)+\frac{1}{3(1+w)}\ln\frac{1-\alpha r}{1-r}\,,\label{eq-e-N2}}
where
\eq{n(\fin_1,\fin_2)\equiv\int_{\fin_2}^{\fin_1}\frac{8\pi\Vi}{\mpl^2\Vi'}\dd\fin\,,}
is the number of e-folds for the part of a single field inflation with $\fin$ varying from $\fin_1$ to $\fin_2$, as if the energy density from $\fcu$ did not exist. The second term in \refeq{e-N2} is the additional number of e-folds provided by the presence of the 
oscillating $\fcu$ or its decay products, however negligible the contribution is. Also
\eq{r\equiv\frac{\Vc\sst}{\Vi\sst+\Vc\sst}\,,}
 and $\alpha\ll1$ has the order of the inflaton's first order slow roll parameter\footnote{If the inflaton $\fin$ remains almost constant we will get $\alpha\rightarrow0$. The exact definition of $\alpha$ is given by:
 \eq{\alpha\equiv1-\frac{24(1+w)\pi\Vi\sst}{\mpl^2}\int_{\fin\sse}^{\fin\sst}\frac{e^{-3(1+w)n(\fin\sst,\fin)}}{\Vi'}\dd\fin\,.}}.

After calculation, the perturbation of $N_2$ from \refeq{e-N2} can be expressed as:
\eq{\delta N_2=-\frac{1-r}{1-\alpha r}\frac{8\pi(\Vi\sst+\Vc\sst)}{\mpl^2\Vc\ssh'}\delta\fcu\ssh,\label{eq-e-dN2}}
which only differs from \refeq{e-dN1} slightly. The dependence on the equation of state $w$ is fully encoded in $\alpha$. The total difference in the number of e-folds of inflation comes from combining \refeq{e-dN1} and \refeq{e-dN2}, during which most terms cancel because of the fluctuating boundary ``$\st$'', yielding:
\eq{\delta N\equiv\delta N_1+\delta N_2=N_{\fcu}\delta\fcu\ssh\label{eq-e-dN}}
where
\eq{N_{\fcu}=\frac{8\pi(1-\alpha)\Vc\sst}{(1-\alpha r)\mpl^2\Vc\ssh'}\approx\frac{8\pi\Vc\ssh}{\mpl^2\Vc\ssh'}\,.\label{eq-e-Np}}

Since $\alpha$ is much smaller than unity and $\Vc\ssh\approx\Vc\sst$, we confirm our previous estimation result in \refeq{e0-dN}. Also, we get the power spectrum of curvature perturbation (according to \refeq{e-Pz0})
\eq{P_\zeta=\frac{16(1-\alpha)^2H\ssh^2\Vc\sst^2}{(1-\alpha r)^2\mpl^4\Vc\ssh'{}^2}\approx\frac{16H\ssh^2\Vc\sst^2}{\mpl^4\Vc\ssh'{}^2}\,.\label{eq-e-Pz}}

From the exact expression of  \refeq{e-Pz}, we also get the exact spectral tilt for the curvature perturbation as:
\eq{n_s-1\equiv\frac{\dd\ln P_\zeta}{H\dd t}=-2\epsilon\ssh+2\eta_{\fcu}\ssh\,,\label{eq-e-ns}}
where $\eta_{\fcu}\ssh\equiv(\mpl^2/8\pi)\Vc\ssh''/(\Vi\ssh+\Vc\ssh)\ll\epsilon\ssh$ is the second order slow roll parameter for $\fcu$.

With the exact expression \refeq{e-Np}, we can now calculate the bispectrum of the curvature perturbations. By taking the derivative $\partial/\partial\fcu\ssh$ on both sides of the exact \refeq{e-Np}, we can define and calculate $N_{\fcu\fcu}\equiv\partial N_{\fcu}/\partial\fcu\ssh$. After some work, we get the strength of the local bispectrum for the curvature perturbation as:
\eq{\fNL=\frac{5}{6}\frac{N_{\fcu\fcu}}{N_{\fcu}^2}=-\frac{5\beta}{6r},\label{eq-e-fnl}}
where $\beta$ is mostly a function of $\alpha$, $r$, $w$ and $\epsilon_{\fin}\sst\equiv\mpl^2\Vi\sst'{}^2/16\pi\Vi\sst^2$ the slow roll parameter for the inflaton itself when $\fcu$ ends slow roll.

Let us now consider a very special case when $\Vc(\fcu)$ is very subdominant so $\Vi(\fin)\gg\Vc(\fcu)$, and $\fcu$ potential is flat enough on the plateau. We can then simplify \refeq{e-Pz} to
\begin{equation}
{\cal P}^{1/2}_{\zeta}\approx\frac{4}{M_p}\left(\frac{H_*}{M_p}\right)\left(\frac{U\ssh}{U\ssh'}\right)\sim 10^{-5}\,,
\end{equation}
and for \refeq{e-fnl}, $\beta$ reduces to
\eq{\beta=\frac{3(1+w)}{1-\alpha}\alpha+2\epsilon_{\fin}\sst\ll1\,.\label{eq-e-betas}}
Since the numerator and the denominator are both much smaller than 1 in \refeq{e-fnl}, whether we get a significant non-Gaussianity then depends on which one is smaller.

\fig{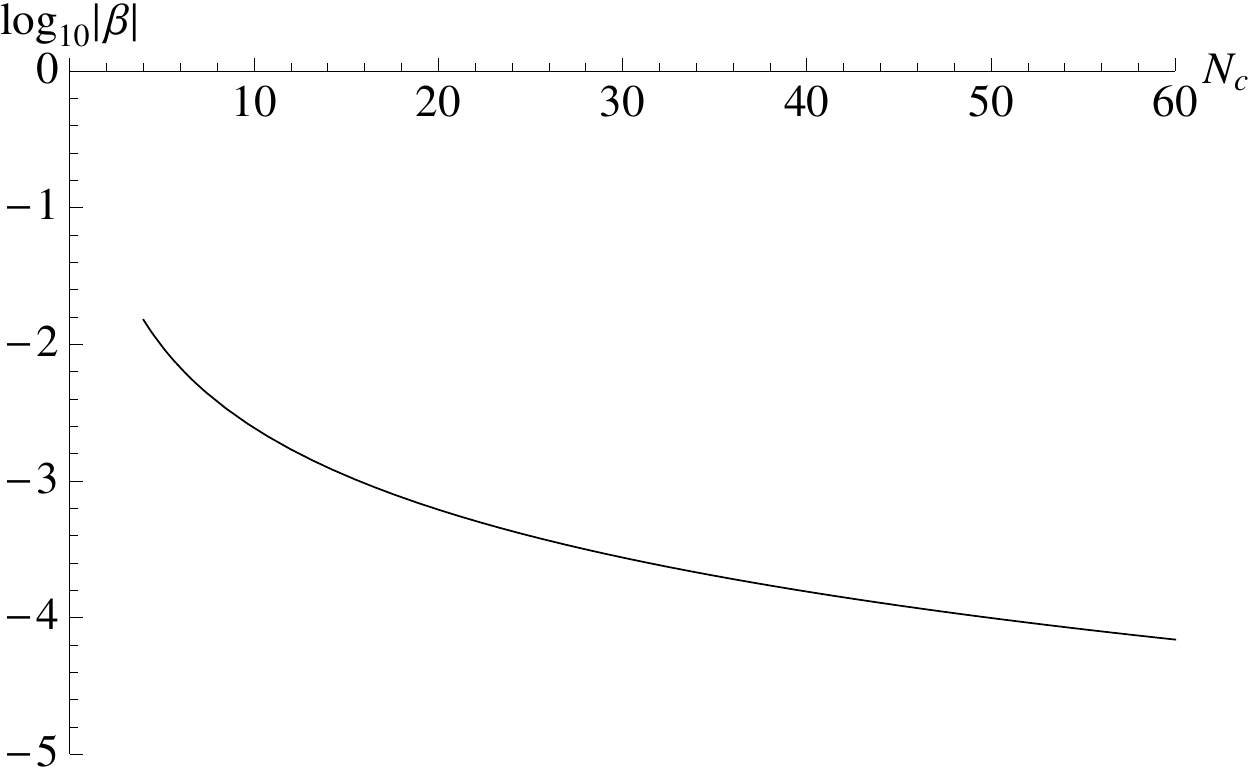}{The $\beta$ parameter for $\Vi(\fin)=m^2\fin^2$ model when $\fcu$ is subdominant. The value of $\beta$ depends on $N\sst$, the remaining number of e-folds of inflation after $\fcu$ ends slow roll. Note here $N=0$ indicates end of inflation.}{0.49}{}
Unlike the power spectrum of curvature perturbations which is determined by $U(\sigma)$ at the Hubble exit, the value of $\beta$ is decided \emph{solely} by the inflaton potential $\Vi(\fin)$ and the equation of state $w$ at the time of phase boundary.  For this reason, the suppression factor $\beta$ only depends on the inflation model. As an illustration, we take $\Vi(\fin)\equiv m^2\fin^2$, and the assumption that $\fcu$ instantly decays into radiation once it ends slow roll. As displayed in \refig{Beta}, we obtain $\beta<0$ while $|\beta|\sim{\cal O}(10^{-4})-{\cal O}(10^{-2})$, depending on $N\sst$, the e-folding when $\fcu$ ends slow roll. 

Furthermore, since the decay products of $\sigma$ dilute away during inflation, there will be no relic isocurvature perturbations. In this respect our paradigm is very different from the curvaton scenario as discussed in Refs.~\cite{curvaton}. One of the challenges for a curvaton paradigm is to ensure that both curvaton and inflaton decay products thermalize which is highly non-trivial. If the curvaton does not dominate the energy density, one has to ensure this happens otherwise large isocurvature perturbations can be produced~\cite{Sesh}. But in this current scenario the origin and the decay products of a decaying field do not lead to any imprint in the thermal history of the universe other than the curvature perturbations.

On another account our model predictions differ from a simple curvaton scenario with a positive curvaton mass~\cite{curvaton}. In this case the curvaton mechanism naturally predicts almost scale invariant perturbations, but in our case we naturally obtain red-tilted spectral index, see \refeq{e-ns}, in accord with the current observations~\cite{WMAP}.

The flatness of $\sigma$ field can be ensured within particle and string theory~\cite{Infl-rev}. However now the onus is to embed $V(\phi)$ within the observable sector such that the inflaton decay products directly excite the Standard Model quarks and leptons. Once this is ensured it is sufficient to create all observed matter and perturbations. In this respect, the decaying field might even originate from the ubiquitous hidden sectors available in beyond the Standard Model physics.

Before we conclude our results, let us briefly mention that the decaying scalar field during inflation can produce significantly large $\fNL$ depending on the value of $r$. For example if $r\approx 10^{-4}$ for the case when $V(\phi)=m^2\phi^2$ with $|\beta|\sim {\cal O}(10^{-3})$, the non-Gaussianity parameter could be of order $\fNL\sim {\cal O}(10)$. If some modes leave the Hubble patch right before $\fcu$ ends slow roll, the $\fNL$ could be made even larger since the suppression factor $\beta\sim{\cal O}(1)$, instead of the suppression given by \refeq{e-betas}. However the spectral tilt could also be significant enough to be ruled out by the current observations. Some of these issues can be investigated in future publications.

Let us conclude by mentioning that we have provided a simple paradigm where a decaying scalar field during inflation can indeed seed red-tilted spectrum with the desired amplitude of perturbations which can give rise to small or large $\fNL$, depending on the ratio $r$ and the potential of the inflaton at the time when the decaying field leaves the slow-roll approximation. This paradigm sources curvature perturbations in a way that has never been discussed before and does not generate isocurvature perturbations.\\\\

AM is supported by the Lancaster-Manchester-Sheffield Consortium for Fundamental Physics under STFC grant ST/J000418/1.

\end{document}